# Preparation for future active learning

Eric W. Burkholder, Mason Sake, Jiamin Zhang
Department of Physics, Auburn University, Auburn AL 36830 USA

It is well documented that students sometimes resist active learning techniques [1]. A recent study showed how students believed that they learned less in active learning classrooms than they learned in lectures, even though they learned more [2, 3]. In this article, we describe a method for introducing active learning methods to college students that is based on a preparation for future learning approach [4]. The students who received this introduction to active learning appeared to be more receptive to group work in the classroom than students who started the course with an explanation of the reasons and values of active learning.

The target course was introductory mechanics for scientists and engineers (calculus-based) at a large public university. It was taught using a productive failure-style (or problem-based learning) active learning approach course in a large lecture hall. Students struggle with problems in small groups before they are told how to solve the problems by the instructor [4, 5]. In the first year of teaching this class, during the first class period, we showed them data supporting the idea that active instruction is more effective than lecture and explained expectations. We used most, if not all, of the strategies identified in Ref. 1. Despite this, there was strong and active resistance to this style of teaching. Some students felt empowered to voice their unhappiness in the middle of class. More often, students complained on course evaluations that they had to "teach themselves" in this style of teaching, or that it was unreasonable to try and work on a problem before they had been told how to solve it.

During the second iteration of the same course, we decided to try and address students' understanding of and buy-in to active learning differently. We drastically re-designed the first two class sessions to introduce students to active learning based on a "preparation for future learning" approach [4]. For the first 15 minutes of the first class session, we reviewed the course syllabus with students, the course elements and grading, and introduced students to the discipline of physics more broadly. We did not explain how the course would be taught. After 15 minutes, we gave them a static equilibrium activity [6] that was designed to get students to invent the ideas of Newton's $1^{st}$ and $3^{rd}$ laws, as well as the idea of contact forces. At the end of this class period, we asked students to write down (and submit) one thing they liked about the class, and one thing they would change. Students generally liked that they were working together in groups. They were less happy that ideas were introduced before we explained the concepts to them.

During the second course section, we began with a review of students' comments to show them that we had noted their concerns. We then asked them to predict which mode of learning they thought was most effective – traditional lectures, active learning in which there was a short lecture and then practice problems, or preparation for future learning. They overwhelmingly thought that active learning was more effective than lecture but thought that having the ideas explained to them first would be more effective. At this point we had created a time for telling [7] and showed them the data on how any active learning is more effective than lecture, but also that it is more effective to struggle with the problem before they are told how to solve it [4, 5]. We then spent a few minutes discussing memory and how the brain works to show them how these active methods strengthened neuronal connections in their brain. We also acknowledged that this mode of learning is difficult and that students typically felt like they were learning less. The remainder of the course was essentially identical to the first iteration.

Throughout the rest of the second iteration of the course, we surveyed students weekly about what was and was not going well in the class. Group work continued to be a highlight for students. Interestingly, we saw almost no comments about students having to teach themselves ideas – only one student of 54 mentioned this on one of the weekly surveys. Students mostly wanted more practice problems, but otherwise had few suggestions for how to improve the course.

It is possible that the instructor gained a reputation for using active learning after the first year, and so that changed who enrolled in their section in the second year. However, we surveyed the students about why they chose this section of the class and over 90% of students gave the time of day as the reason. Another notable change is that we transitioned from a lecture hall in the first semester to an active learning classroom in the second semester, which made it easier for students to talk to one another. However, there are also instances in which cooperative group problem-solving has been shown to work well in large lecture halls [8]. We did take more care the second semester to monitor the groups through surveys and rearrange groups that were not working well together, which may have also contributed to students' overall satisfaction, though typically complaints in the first semester were not related to the students' groups.

We applied ideas from educational psychology about introducing new content to students to introducing new teaching methods. This introduction was much more successful than simple explanation as used in the past.